\documentclass[a4paper,11pt]{article}
\pdfoutput=1 % if your are submitting a pdflatex (i.e. if you have
\usepackage{graphicx}
\usepackage{cite}
\usepackage{amsmath}
\usepackage{amsfonts}
\usepackage{amssymb}
\usepackage{color}
\usepackage{tcolorbox}
\usepackage{float}
\usepackage{multirow,multicol}
\usepackage{tabulary}
\usepackage[flushleft]{threeparttable}
\usepackage{pgfplots,subfigure}
\usepackage{xcolor}
\usepackage{tikz}
\usepgflibrary{arrows}
\usetikzlibrary{shapes.callouts}
\tikzset{  
	level/.style   = { thick, },
	connect/.style = { dotted, red   },
	notice/.style  = { draw, rectangle callout, callout relative pointer={#1} },
	label/.style   = { text width=2cm }
}

\usepackage{jheppub} % for details on the use of the package, please
\usepackage[T1]{fontenc} % if needed
\usepackage{bm}
\usepackage{lscape}
\title{\boldmath Scattering Study of Fermions Due to Double Dirac Delta Potential in Quaternionic Relativistic Quantum Mechanics}

\author[a]{Hassan Hassanabadi}
\author[b]{Hadi Sobhani, \note{Corresponding author.}}
\author[b]{Won Sang Chung}

\affiliation[a]{Physics Department, Shahrood University of Technology, Shahrood, Iran}
\affiliation[b]{Young Researchers and Elite Club, Damghan Branch, Islamic Azad University, Damghan, Iran.}
\affiliation[c]{Department of Physics and Research Institute of Natural Science,
	College of Natural Science, Gyeongsang National University, Jinju 52828, Korea}

\emailAdd{hha1349@gmail.com}
\emailAdd{hadisobhani8637@gmail.com}
\emailAdd{mimip44@naver.net}

\abstract{Scattering discussion due to Double Dirac Equation in Quaternionic version of relativistic quantum mechanics has been studied in this paper in details. In such a quantum mechanics Dirac equation in presence vector and scalar potential has been considered. Then a Quaternionic  double Dirac delta potential comes to our considered system which causes to scatter the particles. Scattering states of the particles have been derived as well as reflected and transmission coefficients are calculated.}

\begin{document} 
\maketitle
\flushbottom

\section{Introduction}
Among a lot of mathematical options in physics, Quaternions are a famous option. Quaternion was initiated by Hamilton for the first time \cite{1,2,3,4,5,6}. Quaternions in general can be represented by
\begin{equation}\label{i-1}
\phi  = {\phi _0} + {\phi _1}{e_1} + {\phi _2}{e_2} + {\phi _3}{e_3},
\end{equation}
where ${\phi _l} (l = 0,1,2,3)$ are real coefficients. There are three imaginary unit in Quaternions which have property
\begin{equation}\label{i-2}
{e_a}{e_b} =  - {\delta _{ab}} + {\varepsilon _{abc}}{e_c}.{\rm{   (a,b,c = 1,2,3)}}
\end{equation} 
If we set $i,j,k$ as the imaginary units, we conclude from Eq.(\ref{i-2}) that
\begin{equation}\label{i-3}
ij =  - ji = k,{\rm{   }}jk =  - kj = i,{\rm{   }}ki =  - ik = j.
\end{equation}
Eq.(\ref{i-3}) tells that in general multiplication of tow Quaternions has non-commutative nature, it means $qp \ne pq$.\\ 
The idea for making use of Quaternions in Quantum mechanics has long a shining story. Physicists and mathematician have  tried to found Quantum mechanics on the Quaternions. Finkelstein et al about new kind of quantum mechanics by using Quaternions made an interesting discussion in \cite{6,7} or some works which are about using real and complexified Quaternions as underlying mathematical structure and by adopting a complex geometry by having ability to present a compatible face of Quantum mechanics \cite{8,9} and there are a lot of valuable researches in this topic which are mentioned in Ref \cite{10,11,12,13,14,15,16,18,19,20,21}. All of them tried to show that we can we a quantum mechanics which based on the Quaternionic. In the rest of paper, we are intend to present Quaternionic form of Dirac equation. We want to extend our recent papers \cite{22, epjc, indian} to relativistic form as well as have a study of scattering by double Dirac delta potential. We organized this article as presenting Quaternionic Dirac equation in Sec. \ref{sec-2}. Introducing double Dirac delta potential and detail discussion about it and its effects in Sec. \ref{sec-3}. And after bringing some information about probability current and conservation law in Sec. \ref{sec-4}, the conclusions are appeared at the end.
\section{Dirac Equation in Quaternionic Quantum Mechanics}
\label{sec-2}
As in ordinary relativistic quantum mechanics, to investigate relativistic fermions, we should make use of Dirac equation. In Quaternionic quantum mechanics also there is Dirac equation but it has some differences. Quaternionic version of Dirac Equation in presence of vector and scalar potential can be written with the help of \cite{23} as
\begin{equation}
\label{2-1}
\frac{{\partial \Psi ({\bm{r}},t)}}{{\partial t}} =  - \left[ {\bm{ \alpha . \nabla}  + {\bm{\beta }}\left( {i(m + {S_a}({\bm{r}})) + j{S_b}({\bm{r}})} \right) + i{V_a}({\bm{r}}) + j{V_b}({\bm{r}})} \right]\Psi ({\bm{r}},t),
\end{equation}
where in Eq. \eqref{2-1} $\hbar  = c = 1$ and
\begin{equation*}
	%\label{2-3}
	\begin{array}{l}
		\bm{ \alpha}  = \left( {\begin{array}{*{20}{c}}
				\bm{ 0} &{\bm{ \sigma} }\\
				{\bm{ \sigma}  }& \bm{ 0} 
			\end{array}} \right)\\
			\\
			{\bm{\beta }} = \left( {\begin{array}{*{20}{c}}
					{\bm{1}}&{\bm{0}}\\
					{\bm{0}}&{{\bm{ - 1}}}
				\end{array}} \right)
			\end{array}
		\end{equation*}
		where Pauli matrices are $\sigma$. As is obviously clear the potentials have two parts. Real functions of them are shown by subscribe $a$(${S_a}({\bm{r}}),{V_a}({\bm{r}}) \in $) and for complex functions, they have subscribe$b$(${S_b}({\bm{r}}),{V_b}({\bm{r}}) \in $). By setting ${S_b}({\bm{r}}),{V_b}({\bm{r}}) \to 0$ we can get to the well-known form of this equation as \cite{25,26}
		\begin{equation}\label{2}
		i\frac{{\partial \Psi ({\bm{r}},t)}}{{\partial t}} = \left( {\bm{ \alpha . P} + {\bm{\beta }}\left( {m + {S_a}({\bm{r}})} \right) + {V_a}({\bm{r}})} \right)\Psi ({\bm{r}},t).
		\end{equation}
		Since we are interested in time-independent interactions, it is better to consider wave function as
		\begin{equation}\label{3}
		\Psi ({\bm{r}},t) = \Phi ({\bm{r}}){e^{ - iEt}}.
		\end{equation}
		Inserting Eq. (\ref{3}) into Eq. \eqref{2-1}, time-independent form of Quaternionic Dirac equation derives
		\begin{equation}\label{4}
		\Phi ({\bm{r}})iE = \left( {\bm{ \alpha . \nabla}  + {\bm{\beta }}\left( {i(m + {S_a}({\bm{r}})) + j{S_b}({\bm{r}})} \right) + i{V_a}({\bm{r}}) + j{V_b}({\bm{r}})} \right)\Phi ({\bm{r}}).
		\end{equation}
		Note the coordinate part of wave function is a Quaternionic function and has components. Actually it usually is written like
		\begin{center}
			$\Phi ({\bm{r}}) = \left( \begin{array}{l}
			{\Phi ^ + }({\bm{r}})\\
			{\Phi ^ - }({\bm{r}})
			\end{array} \right).$
		\end{center}
		
		This form of representing is called spinor form of wave function. If we set $i{S_a}({\bm{r}}) + j{S_b}({\bm{r}}) = i{V_a}({\bm{r}}) + j{V_b}({\bm{r}})$, by some algebraic calculation we arrive at a coupled system of equations for the components
		\begin{align}
			\label{2-5}
			&{\Phi ^ + }(x)iE = {\sigma _x}\frac{{d{\Phi ^ - }(x)}}{{dx}} + \left( {im + 2(i{S_a}(x) + j{S_b}(x))} \right){\Phi ^ + }(x),\\
			\label{2-6}
			&{\Phi ^ - }(x)iE = {\sigma _x}\frac{{d{\Phi ^ + }(x)}}{{dx}} - im{\Phi ^ - }(x).
		\end{align} 
		Quaternionic wave function components are in the form of $\Phi^{\pm} (x) = \phi^{\pm} _a (x) + j \phi^{\pm} _b (x)$ where $\phi^{\pm} _a (x)$ and $ \phi^{\pm} _b (x)$ are the complex functions. Considering such a form of the wave function components in Eq. \eqref{2-6} yields
		\begin{align}
			\label{2-7}
			&\phi^{-} _a (x) = \frac{\sigma_x}{i(E+m)} \frac{d \phi^{+} _a (x)}{dx}, \\
			\label{2-8}
			&\phi^{-} _b (x) = \frac{\sigma_x}{i(E-m)} \frac{d \phi^{+} _b (x)}{dx}.
		\end{align}
		Substitution of Eqs. \eqref{2-7} and \eqref{2-8} into Eq. \eqref{2-5} provides a system of coupled differential equations
		\begin{align}
			\label{2-9}
			&\frac{d^2 \phi^{+} _a (x)}{dx^2} + \left( p^2 -2(E+m)S_a (x)\right) \phi^{+} _a (x) -2i (E+m) S^\ast _b (x) \phi^{+} _b (x) = 0, \\
			\label{2-10}
			&\frac{d^2 \phi^{+} _b (x)}{dx^2} + \left( -p^2 -2(E-m)S_a (x)\right) \phi^{+} _b (x) +2i (E-m) S_b (x) \phi^{+} _a (x) = 0,
		\end{align}
		where $p^2 = E^2 - m^2$ and $\ast$ means the complex conjugation. Now, we are in a position to investigate scattering states due to Quaternionic Dirac delta potential.
		\section{Quaternionic Double Dirac Delta and Scattering}
		\label{sec-3}
		Here, we introduce a Quaternionic form of Double Dirac Delta interaction \cite{indian} as
		\begin{align}
			\label{3-1}
			{S_a}(x) &= {V_a}\left( {\delta (x - a_0) + \delta (x + a_0)} \right)\\ 
			\label{3-2}
			{S_b}(x) &=  i{V_b}\left( {\delta (x - a_0) + \delta (x + a_0)} \right).
		\end{align}
		We assume that $V_a$ and $V_b$ are real constants. The well-known property of Dirac delta interactions is producing discontinuity condition for derivative of wave function. These conditions can be derived by integrating  Eqs. \eqref{2-9} and \eqref{2-10} around $x = a_0$ and $x = -a_0$. So discontinuity condition at $x = a$ is
		\begin{align}
			\label{3-3}
			&\left. \frac{d \phi^{+} _a}{dx}  \right|_{x=a_0 ^+} - \left. \frac{d \phi^{+} _a}{dx} \right|_{x=a_0 ^-} = 2(E+m) (V_a \phi^{+} _a (a_0) + V_b \phi^{+} _b (a_0)), \\ 
			\label{3-4}
			&\left.\frac{d \phi^{+} _b}{dx} \right|_{x=a_0^+}   - \left. \frac{d \phi^{+} _b}{dx} \right|_{x=a_0 ^-} = 2(E-m) ( V_a \phi^{+} _a (a_0) + V_b \phi^{+} _b (a_0)).
		\end{align} 
		as well as for $x = -a$, we have
		\begin{align}
			\label{3-5}
			&\left. \frac{d \phi^{+} _a}{dx}  \right|_{x=-a_0 ^+} - \left. \frac{d \phi^{+} _a}{dx} \right|_{x=-a_0 ^-} = 2(E+m) (V_a \phi^{+} _a (-a_0) + V_b \phi^{+} _b (-a_0)), \\ 
			\label{3-6}
			&\left.\frac{d \phi^{+} _b}{dx} \right|_{x=-a_0^+}   - \left. \frac{d \phi^{+} _b}{dx} \right|_{x=-a_0 ^-} = 2(E-m) ( V_a \phi^{+} _a (-a_0) + V_b \phi^{+} _b (-a_0)).
		\end{align} 
		By the way, we should make three parts in our problem
		\begin{align*}
			&\text{Region I}  &x<-a_0, \\
			&\text{Region II}  &-a_0 <x<a_0, \\
			&\text{Region I}  &x>a_0. 
		\end{align*}
		
		For the free particles we have
		\begin{align}
			\label{3-7}
			&\phi^{+} _a (x) = c_1 e^{ipx} + c_2 e^{-ipx},\\
			\label{3-8}
			&\phi^{+} _b (x) = c_3 e^{px} + c_4 e^{-px}.
		\end{align}
		where the coefficients are complex constants in general.  Therefore, according to our assumption about the particles, the physical wave functions can be written as
		\begin{align}
			\label{3-9}
			&\phi^{+} _{I} (x) =  e^{ipx} + r e^{-ipx} + j e^{px},\\
			\label{3-10}
			&\Phi _{II}^ + (x) = c_1 {e^{ipx}} + c_2 {e^{-ipx}} + j(c_3 {e^{px}} + c_4 {e^{-px}}),\\
			\label{3-11}
			&\phi^{+} _{III} (x) =  t e^{ipx}  + j \tilde{t} e^{-px},
		\end{align}
		
		In order to have explicit expression of the coefficients in Eqs. \eqref{3-9}, \eqref{3-10} and \eqref{3-11} we should match the wave functions at $x = a_0$ and $x = -a_0$ which yield
		\begin{align}
			\label{3-12}
			x = a_0&
			\begin{cases}
				c_1 e^{i a_0 p}+c_2 e^{-i a_0 p}=t e^{i a_0 p}, \\
				c_3 e^{a_0 p}+c_4 e^{-a_0 p}=\tilde{t} e^{-a_0 p},
			\end{cases}
			\\
			\label{3-13}
			x = -a_0&
			\begin{cases}
				r e^{i a_0 p}+e^{-i a_0 p}=c_1 e^{-i a_0 p}+c_2 e^{i a_0 p}, \\
				r e^{i a_0 p}+e^{-i a_0 p}=c_1 e^{-i a_0 p}+c_2 e^{i a_0 p},
			\end{cases}
		\end{align}
		and four equations also are derived by applying continuity conditions 
		\begin{align}
			\label{3-14}
			x=a_0&
			\begin{cases}
				-i c_1 p e^{i a_0 p}+i c_2 p e^{-i a_0 p}+i p t e^{i a_0 p}=2 (E+m)
				\left(V_b \tilde{t} e^{-a_0 p}+t V_a e^{i a_0 p}\right)\\
				p \tilde{t} \left(-e^{-a_0 p}\right)-c_3 p e^{a_0 p}+c_4 p e^{-a_0 p}=2 (E-m)
				\left(V_b \tilde{t} e^{-a_0 p}+t V_a e^{i a_0 p}\right)
			\end{cases}
			\\
			\label{3-15}
			x=-a_0&
			\begin{cases}
				i c_1 p e^{-i a_0 p}-i c_2 p e^{i a_0 p}+i p r e^{i a_0 p}-i p e^{-i a_0 p}=2 (E+m)
				\left(V_b \tilde{r} e^{-a_0 p}+V_a \left(r e^{i a_0 p}+e^{-i a_0
					p}\right)\right)\\
				p \tilde{r} \left(-e^{-a_0 p}\right)+c_3 p e^{-a_0 p}-c_4 p e^{a_0 p}=2 (E-m)
				\left(V_b \tilde{r} e^{-a_0 p}+V_a \left(r e^{i a_0 p}+e^{-i a_0	p}\right)\right)
			\end{cases}
		\end{align}
		So  we have eight equations and un-determined coefficients. By solving these equations, explicit form of each coefficient can be determined but because the solution of these equations were too large we could not be able to bring them.
		
		\section{Probability Current and Conservation Law}
		\label{sec-4}
		
\begin{figure}[H]
			\centering
			\includegraphics[scale=0.4]{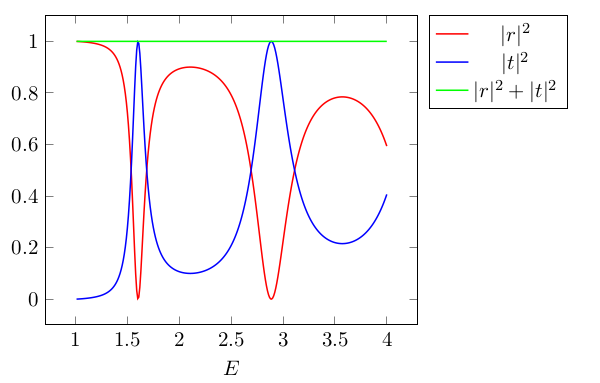}
			\caption{treatments of the coefficients in terms of energy. For this plot we have set $m= a_0 = {V_a } = {V_b } = 1$ and energy$\in [1,4]$. }
			\label{fig 1}
		\end{figure}

\begin{figure}[H]
			\centering
			\includegraphics[scale=0.35]{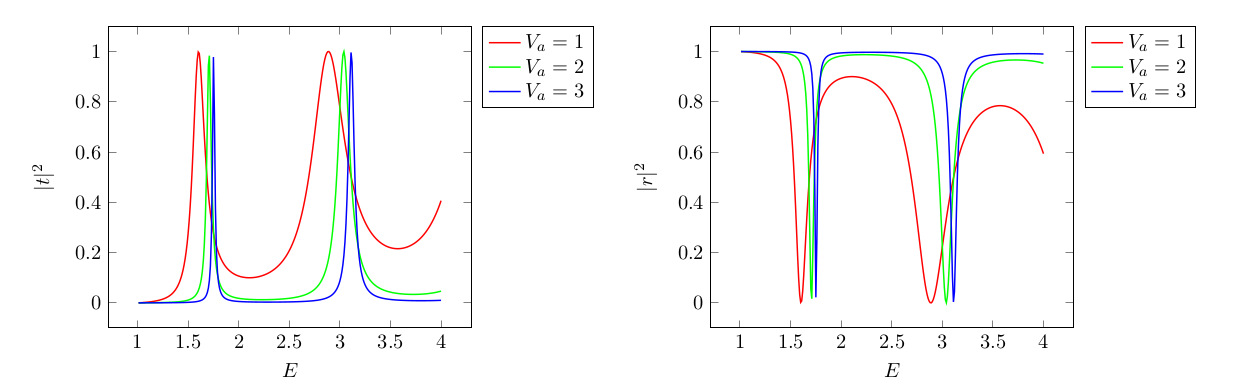}
			\caption{Effects of different values of $V_a$ on the coefficients.}
			\label{fig 2}
		\end{figure}

\begin{figure}[H]
			\centering
			\includegraphics[scale=0.35]{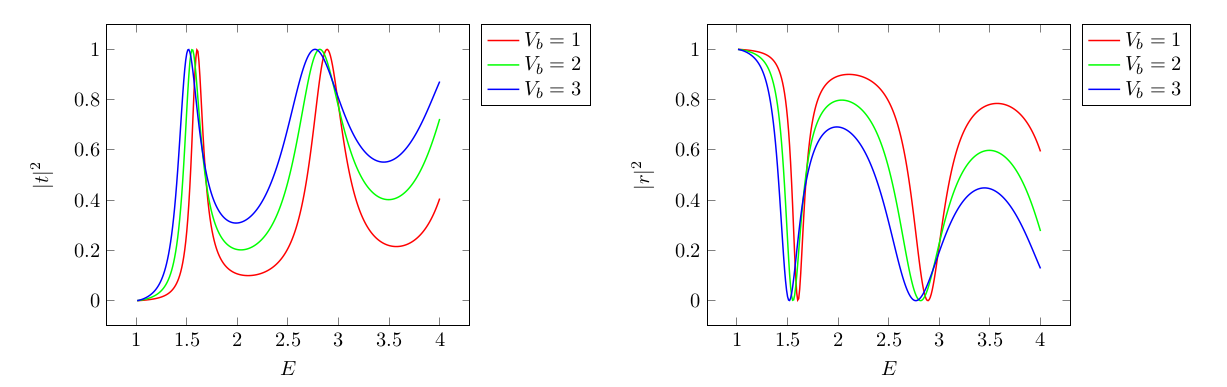}
			\caption{Effects of different values of $V_b$ on the coefficients.}
			\label{fig 3}
		\end{figure}

		In Quaternionic Quantum Mechanics, similar to the complex version, we have the continuity equation as
		\begin{equation}
		\label{4-1}
		\frac{{\partial \rho }}{{\partial t}} + \bm{\nabla.J} = 0,
		\end{equation}
		where
		\begin{align}
			\label{4-2}
			\rho  &= {\bar{\Psi}  }\Psi, \\
			\label{4-3}
			{\bm{J}} &= {\Psi ^\dag } \bm{\alpha} \Psi.
		\end{align}
		To check the conservation law of probability we need to calculate the currents of each regions. To derive the currents of each region, we need to the spinor form of the wave function of each region. This form of the wave functions can be obtained using Eqs. \eqref{2-7} and \eqref{2-8} as
		\begin{align}
			\label{4-4}
			\Phi_I (x) &= 
			\begin{pmatrix}
				e^{ipx} + r e^{-ipx} + j \tilde{r} e^{px}\\
				\sigma_x p \left( \frac{e^{ipx} - r e^{-ipx}}{E+m} + j \frac{\tilde{r} e^{px}}{i(E-m)} \right) 
			\end{pmatrix},\\
			\label{4-5}
			\Phi_{II} (x) &= 
			\begin{pmatrix}
				t e^{ipx}  + j \tilde{t} e^{-px}\\
				\sigma_x p \left( \frac{t e^{ipx}}{E+m} - j \frac{\tilde{t} e^{-px}}{i(E-m)} \right) 
			\end{pmatrix}.
		\end{align}
		It is straightforward to prove that by using the definition $J_x = \Psi^\dagger \alpha_x \Psi$, we derive the constraint
		\begin{align}
			\label{4-6}
			|r|^2 + |t|^2 = 1.
		\end{align}
		So form of conservation law of probability is Eq.(\ref{4-6}). This equation has been plotted in Fig. \ref{fig 1} considering $m= a_0 = {V_a } = {V_b } = 1$ and $E \in [1,4]$.
		As is shown in Fig. \ref{fig 1}, Eq. \eqref{4-6} is valid. It is constructive if we check effects of potential coefficients and distance of the Dirac delta on the reflection and transmission coefficients. In Fig \ref{fig 2} and \ref{fig 3}, effects of potential coefficients on the reflection and transmission coefficients.  It is seen that in Fig. \ref{fig 1} that by increasing $V_a$, the reflection and transmission coefficient appears sharper but in Fig. \ref{fig 2} we face with a different case. When $V_b$ grows up, the reflection and transmission coefficients arises smoother than the in Fig. \ref{fig 2}.
		
		\begin{figure}[H]
			\centering
			\includegraphics[scale=0.35]{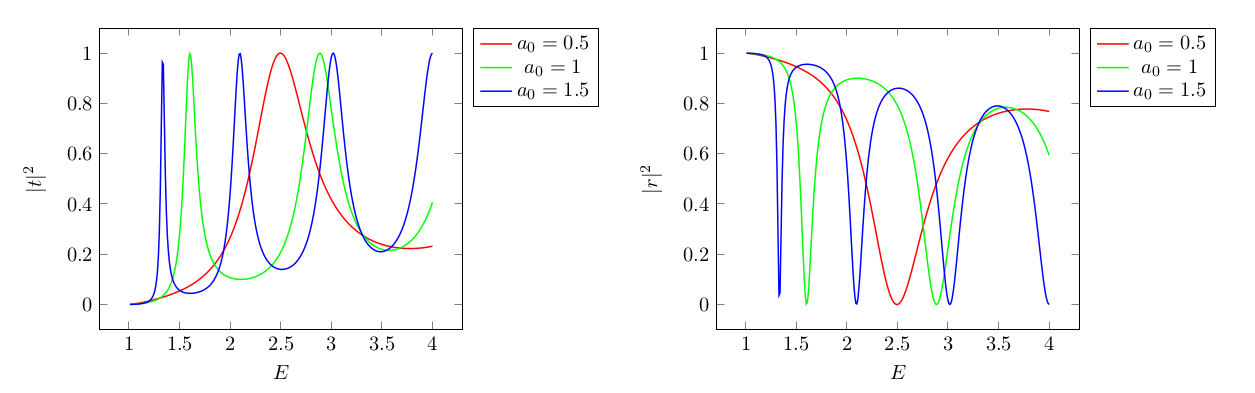}
			\caption{Effects of different values of $a_0$ on the coefficients.}
			\label{fig 4}
		\end{figure}
		In Fig. \ref{fig 4}, by considering  fix values for the potential coefficients and the energy, we change the distance between the double Dirac delta functions. It can be seen that by increasing the distance, the number of fluctuations in the reflection and transmissions increase.
		
		\section{Conclusions}
		
		In this paper, we presented Quaternionic version of Dirac equation in presence of vector and scalar potential. To ensure for correctness of this type of equation, we checked it in special case where we don't have Quaternionic potential, the result was what we expected. after introducing the scattering potential, we investigated effects of it. it causes to the discontinuity conditions for derivative of wave function. probability current for each region of our considered problem was determined. At the end conservation law of probability was derived. It was shown that how different parameters of the scattering potential can affect on the reflection and transmission coefficients for instance  increasing the first part of the potential  causes to have sharper reflection and transmission coefficients but the second part of the potential treated vice versa. It means that by decreasing the second part of the potential we have the sharp reflection and transmission coefficients. For the last case, effects of the distance between the double Dirac delta was shown. It was seen that the number fluctuations of the reflection and transmission increases by enlarging of the distance of the double Dirac Delta.
		
		\appendix
		\begin{landscape}
			\section{Appendix}
			
			In this section details of derivation of Eq. \eqref{4-6} are mentioned. In order to brief the calculation, we will indicate them in a compact form. 
			
			At the first step, we are going to derive  the current of probability of region $I$. Using the definition of the probability current we have
			\begin{align}
				J_{I}=&\Psi^ \dagger \alpha_x \Psi, \nonumber \\
				\label{5-1}
				=&\left( e^{-ipx} + r^\ast e^{ipx} - e^{px} \tilde{r}^\ast j, \quad \sigma_x p \left( \frac{e^{-ipx} - r^\ast e^{ipx}}{E+m} + \frac{\tilde{r}^\ast e^{px}}{i(E-m)}j \right)  \right) 
				\begin{pmatrix}
					0 & \sigma_x \\
					\sigma_x & 0
				\end{pmatrix}
				\begin{pmatrix}
					e^{ipx} + r e^{-ipx} + j \tilde{r} e^{px}\\
					\sigma_x p \left( \frac{e^{ipx} - r e^{-ipx}}{E+m} + j \frac{\tilde{r} e^{px}}{i(E-m)} \right)
				\end{pmatrix}.
			\end{align}
			This point should be noted that the spinor of wave function is a quaternion and since the coefficients are complex constants, order of them and $j$ the imaginary unit is important. Hence in daggered form of the spinor we face with a reversed order in Eq \eqref{5-1}. Proceeding more in the matrix multiplication we have
			\begin{align}
				\label{5-2}
				J_{I}
				=&\left( \overbrace{e^{-ipx} + r^\ast e^{ipx}}^{A_1} \overbrace{- e^{px} \tilde{r}^\ast j}^{A_2}, \quad \sigma_x p \left( \underbrace{\frac{e^{-ipx} - r^\ast e^{ipx}}{E+m}}_{A_3} \underbrace{+ \frac{\tilde{r}^\ast e^{px}}{i(E-m)}j}_{A_4} \right)  \right)
				\begin{pmatrix}
					p \left( \overbrace{\frac{e^{ipx} - r e^{-ipx}}{E+m}}^{B_1} + \overbrace{j \frac{\tilde{r} e^{px}} {i(E-m)}}^{B_2} \right) \\
					\sigma_x\left( \underbrace{e^{ipx} + r e^{-ipx}}_{B_3} + \underbrace{j \tilde{r} e^{px}}_{B_4} \right)  
				\end{pmatrix},\\
				\label{5-3}
				J_{I} =& A_1 B_1 + A_1 B_2 + A_2 B_1 + A_2 B_2 + A_3 B_3 + A_3 B_4 + A_4 B_3 + A_4 B_4.
			\end{align}
			To avoid complicity of multiplication of Eq. \eqref{5-2}, we have 
			
		\end{landscape}
		\begin{subequations}
			\begin{align}
				\label{5-4}
				A_1 B_1 &= \frac{p}{E+m} \left( 1- r e^{-2ipx} + r^\ast e^{2ipx} - |r|^2 \right), \\
				\label{5-5}
				A_1 B_2 &=j \frac{p}{i(E-m)} \left( \tilde{r} e^{px(1+i)} + r \tilde{r} e^{px (1-i)} \right), \\ 
				\label{5-6}
				A_2 B_1 &= j \frac{p}{(E+m)} \left( -\tilde{r} e^{px(1+i)} + r \tilde{r} e^{px (1-i)} \right), \\
				\label{5-7}
				A_2 B_2 &=  \frac{p}{i(E-m)} \left( |\tilde{r}|^2 e^{2px} \right), \\
				\label{5-8}
				A_3 B_3 &=  \frac{p}{E+m} \left( 1 + r e^{-2ipx} - r^\ast e^{2ipx} - |r|^2 \right), \\
				\label{5-9}
				A_3 B_4 &= j \frac{p}{(E+m)} \left( \tilde{r} e^{px(1+i)} - r \tilde{r} e^{px (1-i)} \right), \\
				\label{5-10}
				A_4 B_3 &= -j \frac{p}{i(E-m)} \left( \tilde{r} e^{px(1+i)} + r \tilde{r} e^{px (1-i)} \right), \\
				\label{5-11}
				A_4 B_4 &= -\frac{p}{i(E-m)} \left( |\tilde{r}|^2 e^{2px} \right).
			\end{align}
		\end{subequations}
		With the help of Eqs.\eqref{5-3}-\eqref{5-11}, we can find the  probability current of the region $I$ as
		\begin{align}
			\label{5-12}
			J_{I} = \frac{2p}{E+m} (1-|r|^2).
		\end{align}
		In the same manner, we can find the probability current of region $II$ as
		\begin{align}
			\label{5-13}
			J_{II} = \frac{2p}{E+m} |t|^2.
		\end{align}
		Since there is no sink or source for the particles, we have
		\begin{align}
			\label{5-14}
			J_{I} = J_{II} \Rightarrow |r|^2 + |t|^2 =1. 
		\end{align}

%\appendix
%\section{Some title}
%Please always give a title also for appendices.

%\acknowledgments

%This is the most common positions for acknowledgments. A macro is
%available to maintain the same layout and spelling of the heading.

%\paragraph{Note added.} This is also a good position for notes added
%after the paper has been written.



\begin{thebibliography}{99}

%\bibitem{a}
%Author, \emph{Title}, \emph{J. Abbrev.} {\bf vol} (year) pg.

%\bibitem{b}
%Author, \emph{Title},
%arxiv:1234.5678.

%\bibitem{c}
%Author, \emph{Title},
%Publisher (year).

\bibitem {1}
W. R. Hamilton, \emph{Elements of Quaternions} New York: Chelsea (1969).

\bibitem {2}
W. R. Hamilton, \emph{The Mathematical Papers of Sir William Rowan Hamilton}  Cambridge: Cambridge University Press  (1967).

\bibitem {3} 
A. A. Albert, \emph{Ann. of Math.} {\bf 43} (1942) 161.

\bibitem {4}
B. A. Rosenfeld,  \emph{A History of Non-Euclidean Geometry} Springer-Verlag  (1988).

\bibitem {5}
Carmondy, Kevin,  \emph{App. Math.  Comp} {\bf 84} (1) (1997) 27. 

\bibitem {6}
D. Finkelstein, J. M. Jauch, S. Schiminovich and D. Speiser, \emph{J. Math. Phys.} {\bf 3} (1962)  207 ; {\bf 4} (1963) 788.

\bibitem {7}
D. Finkelstein, J. M. Jauch and D. Speiser, \emph{J. Math. Phys.} {\bf 4} (1963) 136.

\bibitem {8} 
J. Rembielin'ski, \emph{J. Phys. A} {\bf 11} (1978) 2323.

\bibitem {9}
L. P. Horwitz and L. C. Biedenharn, \emph{Ann. Phys.} {\bf 157} (1984) 432.

\bibitem {10}
S. De Leo and G. Ducati, \emph{J. Phys. Math.} {\bf 42}  (2001) 2236.

\bibitem {11}
A. J. Davies and B. H. McKellar, \emph{Phys. Rev. A} {\bf 40} (1989) 4209.

\bibitem {12}
A. J. Davies and B. H. McKellar, \emph{Phys. Rev. A} {\bf 46} (1992) 3671.

\bibitem {13}
S. De Leo, G. Ducati and C. Nishi, \emph{J.Phys. A} {\bf 35}  (2002) 5411.

\bibitem {14}
A. Peres, \emph{Phys. Rev. Lett.}  {\bf 42} (1979) 683.

\bibitem {15}
H. Kaiser, E. A. George and S. A. Werner, \emph{Phys. Rev. A} {\bf 29}  (1984) 2276. 

\bibitem {16}
A. G. Klein, \emph{Physica B} {\bf 151} (1988) 44.

\bibitem {18}
P. R. Girard,\emph{Eur.J.Phys.} {\bf 5} (1984) 25.

\bibitem {19}
K. Shoemake, \emph{Comput. Graph.} {\bf 19} (1985) 245. 

\bibitem {20}
S. Altmann, \emph{Rotations, Quaternions, and Double Groups} Claredon, Oxford  (1986). 

\bibitem {21}
M. Gogberashvili, \emph{Eur. Phys. J. C}  {\bf 74} (2014) 3200.  

\bibitem {22}
H. Sobhani and H. Hassanabadi, \emph{Can. J. Phys.}    {\bf 94} (2016) 262. 

\bibitem{epjc}
H. Sobhani, H. Hassanabadi, and W.S. Chung,  \emph{Eur. Phys. J. C}  {\bf 77}  (2017) 425.

\bibitem{indian}
H. Sobhani and H. Hassanabadi, \emph{Indian J Phys} {\bf 91} (10)   (2017) 1205.  

\bibitem{23}
S. De Leo, G. Ducati and S. Giardino, \emph{J. Phys. Math.} {\bf 6} (2015)  1000130. 


\bibitem{25}
H. Sobhani and H. Hassanabadi,  \emph{Commun. Theor. Phys.} {\bf 64} (2015) 263.

\bibitem{26}
H. Sobhani and H. Hassanabadi,  \emph{Commun. Theor. Phys.} {\bf 65}  (2016) 543.



% Please avoid comments such as "For a review'', "For some examples",
% "and references therein" or move them in the text. In general,
% please leave only references in the bibliography and move all
% accessory text in footnotes.

% Also, please have only one work for each \bibitem.


\end{thebibliography}
\end{document}